\documentclass[aps,prl,twocolumn,amsmath,amssymb,preprintnumbers,floatfix]{revtex4-1}
\usepackage{times}
\usepackage{amsfonts}
\usepackage{amsmath, amsthm, amssymb}
\usepackage{color}
\usepackage{bm}
\usepackage{natbib}
\usepackage[caption=false]{subfig}

\usepackage{graphicx}

\renewcommand{\vec}{\pmb}

\usepackage{soul}

\begin{document}

\title{Macroscale non-local transfer of superconducting signatures to a ferromagnet in a cavity} 

\author{Andreas T.G. Janss{\o}nn, Haakon T. Simensen, Akashdeep Kamra, Arne Brataas and Sol H. Jacobsen}

\affiliation{Center for Quantum Spintronics, Department of Physics, Norwegian University of Science and Technology, NO-7491 Trondheim, Norway}

\begin{abstract}
Cavity spintronics recently heralded non-local magnonic signal transfer between magnetic samples. Here we show that by including superconductors in the cavity, we can make use of these principles to bring composite superconductor--ferromagnet systems to the macroscale. We analyze how a superconductor's a.c. conductivity influences the spin dynamics of a spatially separated magnet, and we discuss the potential impact on spintronic applications.
\end{abstract}

\date{\today}

\maketitle


The field of superconducting spintronics has been gathering pace in the last decade as the promise of achieving low dissipation spin and charge transport has been increasingly refined and realised \cite{linder_nphys_15, EschrigPhysTod2011, eschrig_physrep_15}. It relies on the proximity effect, whereby properties of one material can persist in an adjacent thin film. This places a tight nanometer constraint on the operational range in most cases. The most anomalously long-ranged persistence of superconductive signatures is reportedly up to the micrometer-range \cite{keizer_nature_06,anwar_prb_10}. However, in this paper we highlight the untapped potential of composite superconductor--ferromagnet systems to make use of advances in cavitronics, and that photon-mediated processes can enable the detection of centimeter-ranged superconductive signatures. We provide a readily accessible example to establish the proof of concept, and discuss multiple directions for exploration to highlight the potential for innovation in superconducting spintronic applications.

Cavity spintronics, or cavitronics, is an emerging interdisciplinary field in which microwave or optical cavity photon modes can couple to magnons (also called spin waves). Experiments have shown strong coupling of cavity modes to both ferri- and ferromagnets \cite{Huebl_PRL2013,Bourhill_PRB2016}. This is observed as a hybridization of the photon and magnon modes, indicated by avoided crossings/Rabi splitting in the normal mode frequency spectrum. It was recently shown that magnonic interactions between two non-local magnetic samples can be mediated by the cavity modes \cite{Lambert_PRA2016,Rameshti_PRB2018,Johansen_PRL2018}. This means information encoded in the magnitude and phase of the spin waves (\textit{i.e.} spintronic information) can be transmitted non-locally over macroscopic length scales. We explore the question of magnons coupling non-locally to excitations in a superconductor.

Light with frequencies above the superconducting gap breaks Cooper pairs and thus weakens the superconductivity. However, light can also enhance or induce superconductivity \cite{Light_induced_S1,Light_induced_S2,Light_induced_S3}. In-cavity manipulation of a superconductive component might appear restrictive, demanding effective screening of the contact wires while maintaining the quality factor of the cavity, but also this has been achieved experimentally recently \cite{Tabuchi_Science2015}. In that case, researchers succeeded in driving a black box transmon qubit inside a cavity, coupling the oscillations between the two levels of the qubit to the microwave cavity modes. The transmon qubit is engineered by using the nonlinearity of a superconducting Josephson junction to create an effective two-level system, as in circuit quantum electrodynamics (QED) \cite{Wallraff_Nature2004}. Consequently, this qubit--cavity coupling generated excitement about the potential prospect of unifying quantum optics and solid state quantum computing \cite{Osada_PRL2016,LQ_SciAdv2017}.

Qubit-cavity coupling demonstrated the feasibility of screening wiring to a superconducting system inside microwave cavities. However, superconductivity in that case is used as a means to generate a two-level system, \textit{i.e.} realize a qubit, and not as a means to probe and use the superconductive signatures themselves. By combining standard approaches for the electrodynamics of superconductivity, cavity coupling and magnetism dynamics, we will here provide a proof of principle that there is considerable potential to do just that.


\begin{figure}
\includegraphics[width=0.48\textwidth, angle=0,clip]{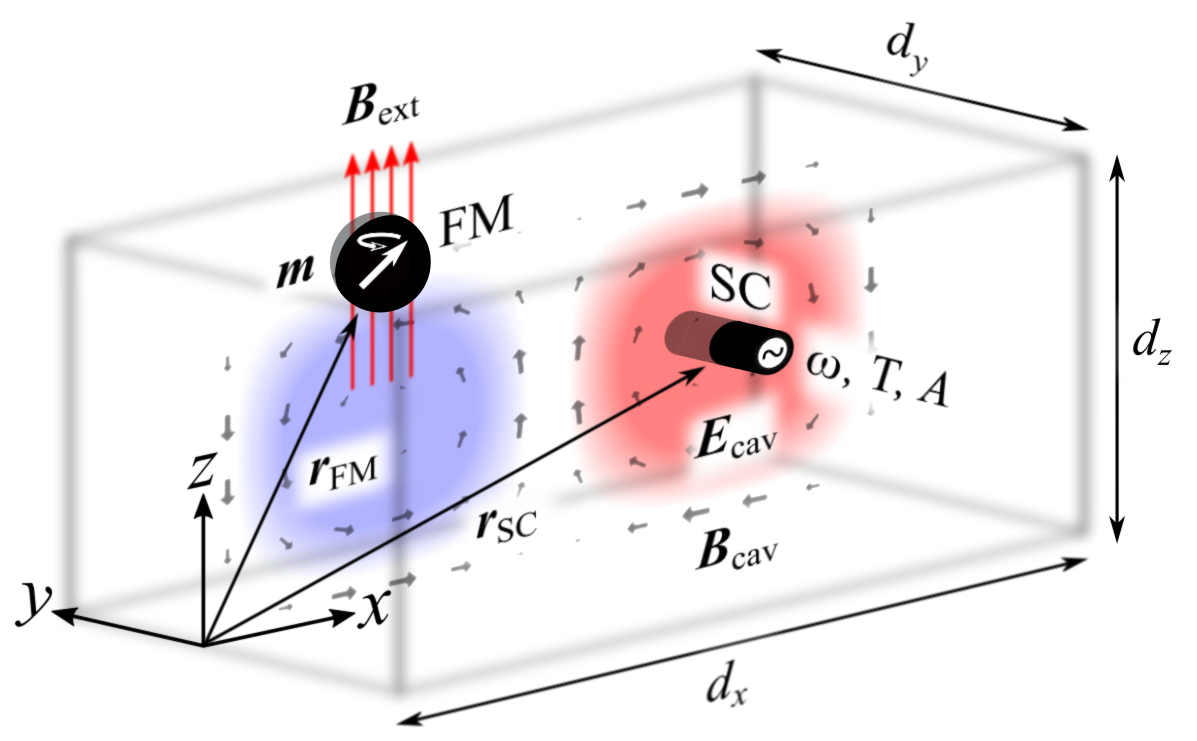}
\caption{(Color online) The proposed model for inducing macroscale photon-mediated superconducting signatures in a magnet (not to scale). The photonic microwave resonator of dimensions $\{ d_x,d_y,d_z \}$ contains a short, thin superconducting wire segment (SC) along the $y$-direction and with a cross-sectional area $A$, connected to an alternating current source via screened wiring through the cavity walls, as well as a small ferromagnetic sphere (FM) with a uniform magnetization $\vec{m}$. The FM and SC are positioned at $\vec{r}_{\mathrm{FM}}$ and $\vec{r}_{\mathrm{SC}}$ respectively, corresponding to extrema of the magnetic and electric components of the cavity mode $\vec{B}_{\mathrm{cav}}$ and $\vec{E}_{\mathrm{cav}}$. Across the SC, $\vec{E}_{\mathrm{cav}}$ is directed along the $y$-axis, and across the FM, $\vec{B}_{\mathrm{cav}}$ is directed along the $x$-axis. The FM is additionally subjected to a strong external magnetostatic field $\vec{B}_{\mathrm{ext}}$ such that $|\vec{B}_{\mathrm{ext}}| \gg |\vec{B}_{\mathrm{cav}}|$, which fixes the precessional axis of $\vec{m}$ along the $z$-direction. We use the TE$_{201}$ cavity mode as an example. The SC current, cavity mode and FM mode couple resonantly at the input a.c. frequency $\omega$. The relative amounts of supercurrent and resistive currents passed through the SC is modulated by the temperature $T$.
}
\label{fig:model}
\end{figure}

We begin by considering the setup illustrated in Fig.~\ref{fig:model}. It depicts a microwave cavity containing an electrically screened thin wire, which has a small exposed superconducting segment (SC) held at temperature $T$, connected to an alternating current (a.c.) source, as well as a small ferromagnetic sphere (FM). The internal current density $\vec{J}$ and electric field $\vec{E}_{\mathrm{SC}}$ of the SC are treated as uniform; \textit{i.e.}, internal spatial variations are neglected. The SC and the FM are placed in regions of maximum electric and magnetic field $\vec{E}_{\mathrm{cav}}$ and $\vec{B}_{\mathrm{cav}}$ of a selected cavity mode, respectively. The dimensions of the SC and the FM are assumed sufficiently small for the local fields across their respective regions to be approximately uniform, and their spatial extension are effectively taken to be line-like and point-like at positions $\vec{r}_{\mathrm{SC}}$ and $\vec{r}_{\mathrm{FM}}$, respectively.

The SC is directed along the $y$-direction, and has a critical temperature $T_c$. The a.c. source produces signal frequency $\omega$, which is resonant with the cavity frequency and the frequency of the precessing FM magnetization. By lowering $T$, we pass through the superconducting transition and induce a change in the superconductor’s conductivity. This in turn alters the excitation of the cavity, and the resultant effect on the spin dynamics in the magnet can be harnessed as a non-local detector. That is, by exploiting the mutually resonant coupling to the cavity, it is possible to probe the superconducting transition via a change in the magnonic precession response. We consider the weak coupling approximation, in which the back-action does not alter the physical response of either system (the back-action cannot alter the established electromagnetic response of the superconducting transition).

\begin{figure}[b]
\includegraphics[width=0.48\textwidth, angle=0,clip]{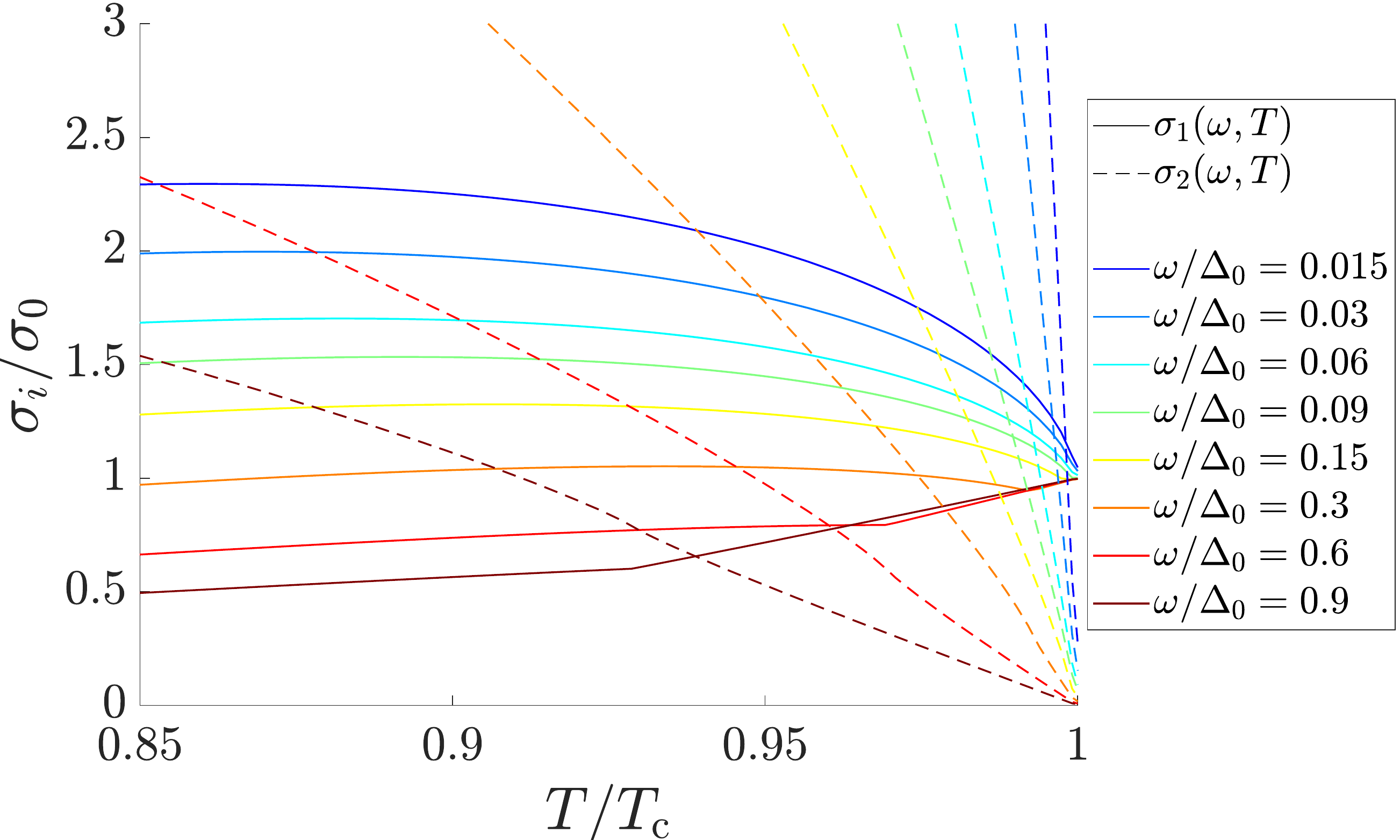}
\caption{(Color online) Intersections of the real ($\sigma_1$) and negative imaginary ($\sigma_2$) part of the SC conductivity $\sigma = \sigma_1 - i\sigma_2$, as a function of $T$, for frequency inputs $\omega$. Material parameters for Nb are used ($T_c = 9.26$~K)  \cite{Matthias1963}; $\sigma_0$ is the normal state direct current conductivity. These plots are generated numerically using Mattis--Bardeen theory \cite{Mattis1958}.
}
\label{fig:conductivities}
\end{figure}

As a concrete example, we consider the TE$_{201}$ cavity mode, where $\vec{E}_{\mathrm{cav}}$ is directed along the $y$-axis over the SC, and $\vec{B}_{\mathrm{cav}}$ along the $x$-axis over the FM. $\vec{B}_{\mathrm{cav}}$ then couples predominantly to the Kittel mode of the FM, \textit{i.e.} the uniform mode of the spherical spin field, quantified by the unit magnetization vector $\vec{m}$. The FM is additionally exposed to a relatively strong external magnetic field $\vec{B}_{\mathrm{ext}}$ such that $|\vec{B}_{\mathrm{ext}}| \gg |\vec{B}_{\mathrm{cav}}|$, which fixes the precessional axis of $\vec{m}$ along the $z$-direction. $|\vec{B}_{\mathrm{ext}}|$ also regulates the resonance frequency of the spin field mode, and reduces the impact of $\vec{B}_{\mathrm{cav}}$ to small perturbations on the motion of $\vec{m}$. The resonance frequency of the TE$_{201}$ mode is determined by $\{ d_x,d_z \}$, which one may thus match to the resonance frequency of the Kittel mode and the frequency of the input a.c. by adjusting $|\vec{B}_{\mathrm{ext}}|$ and $\omega$.

The current response of a superconductor to an applied electric field, taking into account both frequency and temperature, may be derived from microscopic theories of superconductivity, such as BCS- or Eliashberg-theory. Mattis--Bardeen theory is derived from the former~\cite{Mattis1958, Dressel2013}, and provides accurate descriptions of the optical conductivity of BCS superconductors. However, these theories are generally cumbersome to deal with analytically, and will be reserved for numerical calculations. To analytically model the transition from resistive to superconducting current in the SC, we employ the well established framework of the two-fluid model~\cite{Tinkham}.

The SC is treated as two parallel channels carrying normal ($n$) and superconducting ($s$) electrons, respectively. The superconducting channel is characterized by an asymptotically infinite relaxation time $\tau_s \longrightarrow \infty$, and the normal channel assumes a low input frequency $\omega \tau_n  \ll 1$ relative to the relaxation time of $n$-electrons. In this case,
\begin{align}
    \frac{\mathrm{d}\vec{J}_s (\omega, T, t)}{\mathrm{d}t} = \frac{N_s(T) e^2}{m_e}\vec{E}_{\mathrm{SC}}(\omega, T, t), \label{eq:Drudemodels} \\
    \frac{\vec{J}_n (\omega, T, t)}{\tau_n} = \frac{N_n(T) e^2}{m_e}\vec{E}_{\mathrm{SC}}(\omega, T, t), \label{eq:Drudemodeln}
\end{align}
where $m_e$ is the electron mass, and $\vec{J}_i$ and $N_i$ are the current and electron densities of the respective channels. For sinusoidal time dependencies there is therefore a relative phase difference of $\pm\pi/2$ between the contributions of $\vec{J}_s$ and $\vec{J}_n$ to $\vec{E}_{\mathrm{SC}}$ in a current-driven system. $\vec{E}_{\mathrm{SC}}$ thus acquires a phase relative to the net current density $\vec{J} = \vec{J}_n + \vec{J}_s$ between $0$ and $\pm\pi/2$. We argue that this phase shift can be used to bridge superconducting and spintronic circuits via non-local coupling to magnons. In this case it can monitor the superconducting transition, and be implemented as a superconducting switch. More broadly, it opens the door for wider investigations of macro-scale effects in superconducting circuits.

Upon connecting the SC to an a.c. source, the net current density magnitude $J(\omega,t) = I \exp(i \omega t) / A$, where $I$ is the current amplitude, $A$ the SC cross-sectional area, and $\omega$ the input frequency. Inserting into Eqs.~\eqref{eq:Drudemodels} and \eqref{eq:Drudemodeln}, we have
\begin{equation}\label{eq:ESCcurrensource}
    E_{\mathrm{SC}}(\omega,T,t) = \frac{I}{A\sigma(\omega,T)} \exp(i \omega t),
\end{equation}
where
\begin{equation}\label{eq:sigmaSC}
    \sigma(\omega,T) = \frac{e^2}{m_e}\left( N_n(T) \tau_n - i \frac{N_s(T)}{\omega} \right) \equiv \sigma_1(T) - i \sigma_2(\omega,T).
\end{equation}
The phenomenological temperature dependency of $N_i$, and by extension $\sigma_1$ and $\sigma_2$, is 
\begin{equation}
    N_s(T) = N\left[ 1 - \left( T / T_c \right)^4 \right], \quad N_n(T) = N \left( T / T_c \right)^4,
\end{equation}
where $N$ is the total electron density, and $T \leq T_c$ \cite{Tinkham}. For the purpose of analytic insight we retain this simple form, although we include the standard temperature modification of the gap in the numerics \cite{Tinkham}. Above $T_c$, $\sigma$ reduces to the normal metal direct current conductivity $\sigma_0 \equiv N e^2 \tau_n / m_e$. Note that according to Mattis--Bardeen theory, $\sigma_1$ is frequency dependent; near $T_c$, it has a pronounced coherence peak at lower frequencies, and a kink at higher frequencies due to optical excitations across the superconducting gap (see Fig.~\ref{fig:conductivities})~\cite{Mattis1958, Dressel2013}. Neither feature is captured by the two-fluid model. Nevertheless, in terms of the relative magnitudes of $\sigma_1$ and $\sigma_2$, and their point of intersection marking the boundary between the superconducting and resistive regimes, the two-fluid model and Mattis--Bardeen theory coincide very well at the experimentally relevant lower frequencies. Fig.~\ref{fig:conductivities} thus shows the predicted temperatures for the transition between normal and superconducting current \footnote{The high frequency cases are illustrative for materials with lower critical temperature at the relevant lower frequencies.}.

$\vec{E}_{\mathrm{SC}}$ and $\vec{E}_{\mathrm{cav}}$ are assumed to be purely tangential to the SC--cavity interface in our setup (see Fig.~\ref{fig:model}). Thus, by the continuity of the tangential electric field across any interface, $
    \vec{E}_{\mathrm{SC}}(\vec{r}_{\mathrm{SC}},\omega,T,t) = \vec{E}_{\mathrm{cav}}(\vec{r}_{\mathrm{SC}},\omega,T,t)
$ at the surface of the SC. Upon computing the cavity modes by imposing rectangular boundary conditions on the fields, one finds that across the FM and specifically for the TE$_{201}$ mode, $\vec{B}_{\mathrm{cav}}$ at the FM is~\cite{Poole1983} \footnote{The minus sign in the last line of Eq.~(\ref{eq:Bcav0}) is due to the relative positions of the SC and the FM in Fig.~\ref{fig:model}.}
\begin{equation}\label{eq:Bcav0}
    \begin{split}
        \vec{B}_{\mathrm{cav}}(\vec{r}_{\mathrm{FM}},\omega,T,t) = & B_{\mathrm{cav}}(\vec{r}_{\mathrm{FM}},\omega,T,t) \hat{x} \\
        = & -\frac{\pi E_{\mathrm{cav}}(\vec{r}_{\mathrm{SC}},\omega,T,t)}{i \omega d_z} \hat{x}.
    \end{split}
\end{equation}
Furthermore, the resonance frequency of the TE$_{201}$ mode is
\begin{equation}\label{eq:omegacav}
    \begin{split}
        \omega = c \sqrt{\left( \frac{2\pi}{d_x} \right)^2 + \left( \frac{\pi}{d_z} \right)^2},
    \end{split}
\end{equation}
where $c$ is the speed of light in vacuum. With $d_x$ and $d_z$ given, this equality for resonant coupling is ensured by tuning $\omega$. 

The precessional motion of the FM magnetization vector $\vec{m}$ is adequately described by the Landau--Lifshitz--Gilbert (LLG) equation:
\begin{equation}
    \begin{split}\label{eq:LLGeq}
        \frac{\partial \vec{m}\left(\omega, T, t \right)}{\partial t} = & -\gamma \vec{m}\left(\omega, T, t \right) \times \vec{B}\left(\omega,T, t \right) \\
        & + \alpha \vec{m}\left(\omega, T, t \right) \times \frac{\partial \vec{m}\left(\omega, T, t \right)}{\partial t}.
    \end{split}
\end{equation}
Here, $\gamma$ and $\alpha$ are the gyromagnetic ratio and the phenomenological damping parameter of the LLG equation, respectively. $\vec{B}$ is the effective magnetic field inside the FM, including the external, the demagnetization and the magnetocrystalline anisotropy field \cite{Tserkovnyak2002,Risinggaard2017}. The latter two are generally influenced by the geometry and crystal structure of the FM, and may influence $\omega$ and the orbit of $\vec{m}$. We assume an easy axis such as $\langle 111 \rangle$ for YIG \cite{Wu2010}, coinciding with the $z$-direction; and negligible demagnetization and anisotropy fields relative to $\vec{B}_{\mathrm{ext}}$. The latter is reasonably expected to hold down to an input frequency of 5 GHz \cite{Wu2010,Kalappattil2017,Lee2016,Schreier2015}. The effective magnetic field across the FM is then
\begin{equation}\label{eq:Btotal}
    \begin{split}
        \vec{B}(\omega,T,t) & = \vec{B}_{\mathrm{cav}}(\vec{r}_{\mathrm{FM}},\omega,T,t) + B_{\mathrm{ext}} \hat{z} \\
        & = -\frac{\pi E_{\mathrm{SC}}(\vec{r}_{\mathrm{SC}},\omega,T,t)}{i \omega d_z} \hat{x} + B_{\mathrm{ext}} \hat{z}.
    \end{split}
\end{equation}

\begin{figure}[b]
     \centering
     \includegraphics[width=0.48\textwidth, angle=0,clip]{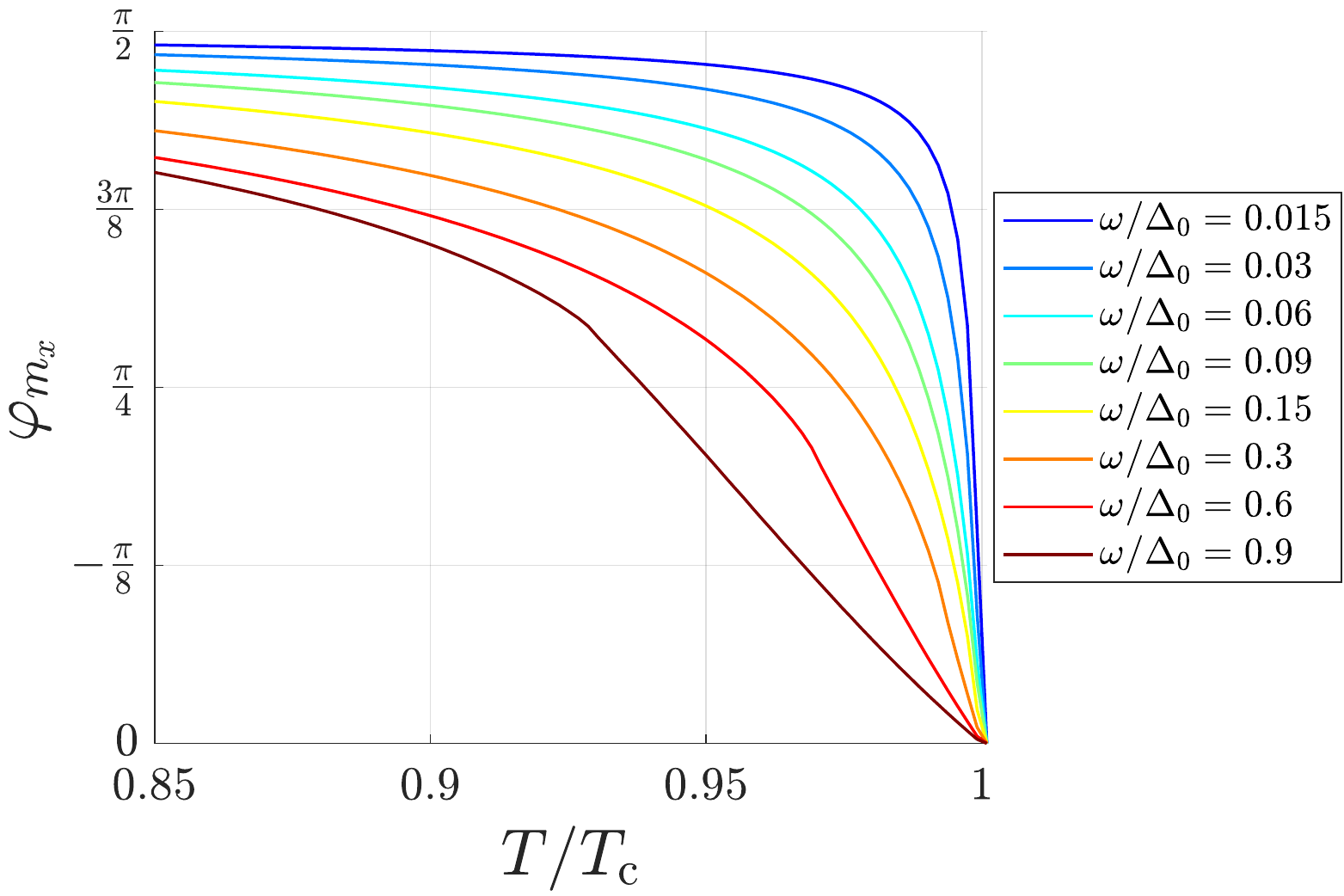}
     \caption{(Color online) Phase of the magnon precession $\varphi_{m_x}$, as a function of $T$, for frequency inputs $\omega$, using Mattis--Bardeen theory to compute the SC conductivity. Material parameters for Nb and YIG are used, with $T_c = 9.26$~K and $\alpha = 10^{-5}$ \cite{Matthias1963,Risinggaard2017,Ferona2017,Flebus2017}. This phase may be measured relative to the input signal passed through the SC, and its value indicates the relative presence of supercurrent and resistive current in the SC. For this $\alpha$, $\varphi_{m_y} \approx \varphi_{m_x} -  \pi/2$.}
     \label{fig:phases}
\end{figure}

When $|B_{\mathrm{ext}}| \gg |\vec{B}_{\mathrm{cav}}|$, $m_z \approx 1 \gg |m_x|, |m_y|$, 
to first order. In Eq.~\eqref{eq:LLGeq}, terms of higher order than linear in $B_{\mathrm{cav}}$, $m_x$ and $m_y$, may then be neglected. In addition, the coupling between the cavity mode and the FM is resonant by design. Solving the LLG equation with complex time dependencies $\exp\left( i \omega t \right)$ in $\vec{B}$ and $\vec{m}$, one finally extracts the real parts as physical solutions  \footnote{In contrast, the general LLG equation mixes real and imaginary parts, so they are generally not solutions to the LLG equation by themselves.}. Note that $B_{\mathrm{cav}}$ oscillates exclusively along the x-axis, which breaks the symmetry of the linearized LLG equation. The resulting orbits are conse-
quently elliptical. The expression for $\vec{m}$ therefore has the form $\vec{m}(\omega,T,t) \approx \hat{z} + \vec{m}_p(\omega,T,t)$,
with precessing component
\begin{equation}\label{eq:mprecessing}
        \vec{m}_p(\omega,T,t) = [ m_x(\omega,T) \hat{x} + m_y(\omega,T) \hat{y} ] \exp(i \omega t).
\end{equation}
\begin{figure}[t]
\includegraphics[width=0.48\textwidth, angle=0,clip]{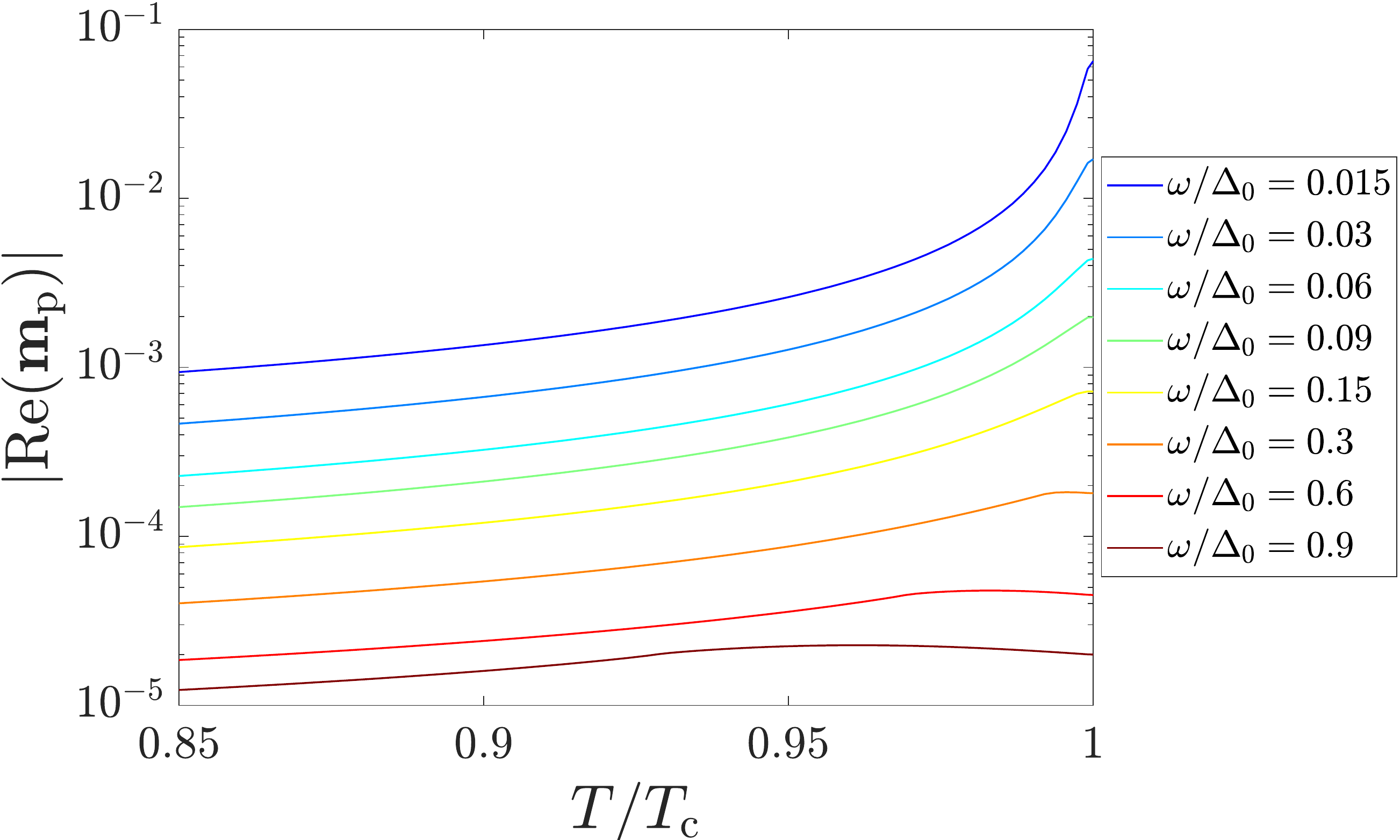}
\caption{(Color online) The magnitude of the precessing component of the magnetization vector $|\mathrm{Re} (\vec{m}_p)|$, as a function of $T$, for frequency inputs $\omega$, using Mattis--Bardeen theory to compute the SC conductivity. Material parameters for Nb, a microwave cavity and YIG are used, with $T_c = 9.26$~K, $\alpha = 10^{-5}$, $\gamma = 176$~GHz/T, $I = 0.6$~A, $A = 10^{-11}$~cm$^2$ and $d_z = 5$~cm \cite{Matthias1963,Risinggaard2017,Ferona2017,Flebus2017}. Within experimental limits such as the critical current of the SC, the decrease in magnitude for increasing frequencies may be counteracted by increasing the input current.}
\label{fig:mprecessing}
\end{figure}

Solving Eq.~\eqref{eq:LLGeq} for $m_x$ and $m_y$ and assuming weak damping $\alpha \ll 1$, one finds the phases relative to the input a.c.~\cite{footnote:supplementary}:
\begin{align}
        \varphi_{m_x}(\omega,T) & \equiv \arg(m_x(\omega,T)) 
        \approx \arctan\frac{\sigma_2(\omega,T)}{\sigma_1(T)} \!+\! \frac{\alpha}{2},\label{eq:argmx}\\
        \varphi_{m_y}(\omega,T) & \equiv \arg(m_y(\omega,T)) 
        \approx \arctan\frac{\sigma_2(\omega,T)}{\sigma_1(T)} \!-\! \frac{\alpha}{2} \!-\! \frac{\pi}{2}.\label{eq:argmy}
\end{align}
Reinserting the solutions for $m_x$ and $m_y$ into Eq.~\eqref{eq:LLGeq}, then taking the absolute value of both sides, yields $\omega = |\gamma B_{\mathrm{ext}}|$. For a given $\omega$, this equality for resonant coupling is ensured by tuning $B_{\mathrm{ext}}$. 

The phase and magnitude of the magnon precession allows us to extract measurable spintronic responses to changes in the superconductor. The magnitude of the precessing component $\left| \mathrm{Re} (\vec{m}_p) \right|$ relates to the cone angle of the precession, and is given by \footnote{$\left| \mathrm{Re} (\vec{m}_p) \right|\lesssim 0.1$ ensures consistency with the approximations of linear response and elliptical orbits. The real part of $\vec{m}_p$ is used here as it is a physical solution of the linearized LLG equation. The complex vector $\vec{m}_p$ is not a physical solution.}
\begin{equation}\label{eq:precessioncomp}
\begin{split}
    & | \mathrm{Re} [\vec{m}_p(\omega,T,t)] | \approx \\
    & | m_y(\omega,T) | \sqrt{2\alpha  \cos^2\left( \omega t \!+\! \varphi_{m_y} \!+\! \alpha \!+\! \frac{\pi}{2} \!-\! \theta \right) \!-\! \alpha \!+\! 1 },
\end{split}
\end{equation}
where
\begin{align}
\begin{split}
    \left| m_y(\omega,T) \right| &\approx \frac{|\gamma| \pi I}{2A |\sigma(\omega,T)| \omega^2 d_z \alpha},
\end{split} \label{eq:absmy} \\ 
\begin{split}
    \theta &\approx \frac{3\pi+\alpha}{4}.
\end{split}\label{eq:theta}
\end{align}
Within experimental limits such as the critical current of the SC, $\left| \mathrm{Re} (\vec{m}_p) \right| \lesssim 0.1$ may easily be achieved by regulating the input current amplitude $I$. Above this value, second and higher order corrections of the orbit become significant, and the full LLG must be employed. Note that for a negligible $\alpha$, $\left| \mathrm{Re} (\vec{m}_p) \right|$ becomes independent of time; the orbit is then circular with $\varphi_{m_y} \longrightarrow \varphi_{m_x} - \pi/2$. 



Plots of $\varphi_{m_x}$ and $|\mathrm{Re} (\vec{m}_p)|$ with realistic parameters using Mattis--Bardeen theory are presented in Figs.~\ref{fig:phases}~and~\ref{fig:mprecessing}. Eqs.~\eqref{eq:argmx} and \eqref{eq:argmy} show as expected that in passing from a superconducting regime, \textit{i.e.} $\sigma_2 \gg \sigma_1$, to a resistive regime, \textit{i.e.} $\sigma_2 \ll \sigma_1$, the phase of $\mathrm{Re} (\vec{m}_p)$ will shift by $-\pi/2$, exactly corresponding to the simultaneous shift in $\vec{E}_{\mathrm{SC}}$. Moreover, it becomes clear from Eqs.~\eqref{eq:argmx}--\eqref{eq:precessioncomp} that as the FM damping $\alpha$ increases, the orbit becomes tilted in the $xy$-plane with respect to its principal axes, and becomes progressively more eccentric~\cite{footnote:supplementary}. The tilting angle between the $x$-axis and the major axis of the elliptic orbit is $\theta$ as given by Eq.~\eqref{eq:theta}. This phenomenon may be of particular interest in future works if one couples the FM and the SC by circularly instead of linearly polarized light, and if one operates with triplet instead of singlet superconductivity.

The above coupling mechanism shows clearly that a transition from the resistive to the superconducting state translates directly to a measurable non-local phase shift in the magnon precession frequency, with an experimentally resolvable perturbation of $\vec{m}$ of a few percent expected to be possible for various choices of magnetic and superconducting materials. The magnon excitations can be incorporated into extended spintronic circuitry outside the cavity, with no proximity coupling to the SC required. The shift in $\varphi_{m_x}$ and $\varphi_{m_y}$ may be measured \textit{e.g.} via Faraday rotation \cite{Xia2015, Kampfrath2010}, or via a.c. spin pumping \cite{Jiao2013,Weiler2014,Hahn2013,Wei2014,Tserkovnyak2002}. The method of Faraday rotation has sufficient resolution to detect single oscillations in the resonance frequency regimes of interest. The phase can then be measured relative to the a.c. input signal, as a function of the input frequency $\omega$. Alternatively, a.c. spin-pumping would be more easily achieved by changing the geometry of the ferromagnetic sphere to a film with deposited platinum layer. The analytics would then require the inclusion of the demagnetization field and associated shift in resonance, but it would not otherwise alter the physics. 


This work shows that photon-mediated superconducting signatures are a feasible way to provide a bridging circuit for spintronic applications. In device design this can feature as a superconductive switch, but also to monitor the superconducting transition and critical temperature of the superconductor directly. 

However, the importance of the result also goes beyond these applications as it opens up a plethora of interesting investigative avenues. For example, by switching from a conventional singlet superconductor to a triplet source (either intrinsically $p$-wave or odd-frequency $s$-wave), then there are no longer two simple coupling relationships to the cavity as in the case of the a.c.-driven oscillators in Eqs.~\eqref{eq:Drudemodels}~and~\eqref{eq:Drudemodeln}. The nature of this coupling remains to be explored, but it seems plausible in that case that one may employ the cavity setup to probe and differentiate between the different current components. This may make cavity spintronics with superconductors -- or super cavitronics -- an interesting new tool for probing unconventional superconductors.

For the physical picture presented above, it is sufficient to consider a classical description of the coupling. However, it would be interesting to explore a microscopic picture along the line of cavity QED as outlined in Ref.~\cite{Cottet_preprint2019}. In that case we can of course not neglect the details of the mesoscopic circuit by tracing over the mesoscopic degrees of freedom, meaning the mathematical approach becomes rather involved. Nevertheless, it is expected to yield valuable insight to the case of fermionic reservoirs in a cavity.

\begin{acknowledgments}
\textit{Acknowledgments}. 
We thank H. Huebl for useful discussions. We acknowledge funding via the ``Outstanding Academic Fellows'' programme at NTNU, the Research Council of Norway Grant number 302315, as well as through its Centres of Excellence funding scheme, project number 262633, “QuSpin”.
\end{acknowledgments}

\bibliographystyle{apsrev4-1}

\clearpage

\onecolumngrid

{\centering
\section*{\Large Supplementary material}}

\twocolumngrid

\section*{Solutions of the Linearized LLG Equation}

The Landau--Lifshitz--Gilbert (LLG) equation reads:
\begin{equation}
    \begin{split}\label{supp:eq:LLGeq}
        \frac{\partial \pmb{m}\left(\omega, T, t \right)}{\partial t} = & -\gamma \pmb{m}\left(\omega, T, t \right) \times \pmb{B}\left(\omega,T, t \right) \\
        & + \alpha \pmb{m}\left(\omega, T, t \right) \times \frac{\partial \pmb{m}\left(\omega, T, t \right)}{\partial t}.
    \end{split}
\end{equation}
Upon assuming that $|B_{\mathrm{ext}}| \gg |\pmb{B}_{\mathrm{cav}}|$, $m_x$ and $m_y$ will also be much smaller than $m_z$. In Eq.~\eqref{supp:eq:LLGeq}, terms of higher order than linear in $B_{\mathrm{cav}}$, or the remaining components $m_x$ and $m_y$, may then be neglected. Furthermore, $m_z$ may be taken to be unity to first order. 
The orbit of $\pmb{m}$ is then confined to the $xy$-plane. Eq.~\eqref{supp:eq:LLGeq} thus reduces to two coupled equations:
\begin{align}
    \partial_t m_x & = - \left( \gamma B_{\mathrm{ext}} + \alpha \partial_t \right) m_y, \\
    \partial_t m_y & = \left( \gamma B_{\mathrm{ext}} + \alpha \partial_t \right) m_x - \gamma B_{\mathrm{cav}}.
\end{align}
These are the components of the linearized LLG equation. Note that the presence of $\pmb{B}_{\mathrm{cav}}$, which oscillates exclusively along the $x$-axis, breaks the symmetry of these equations.

Assuming a resonant coupling to $\pmb{B}_{\mathrm{cav}}$, one may thus anticipate a steady elliptic precession in the $xy$-plane:
\begin{equation}\label{supp:eq:mprecessing}
\begin{split}
        \pmb{m}_p(\omega,T,t) & \equiv m_x(\omega,T,t) \hat{x} + m_y(\omega,T,t) \hat{y} \\
        & = [ m_x(\omega,T) \hat{x} + m_y(\omega,T) \hat{y} ] \exp(i \omega t),
\end{split}
\end{equation}
such that $\pmb{m} = \hat{\pmb{z}} + \pmb{m}_p$. The complex expressions for $m_x$ and $m_y$, derived from the linearized LLG equation, are then
\begin{equation}
    m_y =\frac{m_x}{i-\alpha} = \frac{i(\alpha+2i)B_{\mathrm{cav}}}{\alpha(\alpha^2+4)B_{\mathrm{ext}}},
\end{equation}
where $B_{\mathrm{cav}}$ is
\begin{equation}
    B_{\mathrm{cav}} = -\frac{\pi I}{i A \sigma \omega d_z},
\end{equation}
\textit{cf.} Eq. (3) and (6) in the manuscript. For $\alpha \ll 1$,
\begin{equation}
    m_x = -im_y = -\frac{B_{\mathrm{cav}}}{2B_{\mathrm{ext}}\alpha}.
\end{equation}
Furthermore, $|B_{\mathrm{ext}}| = \omega/|\gamma|$ by assumption of resonance. Eq. (14) in the manuscript follows from the above.

\onecolumngrid

\section*{Illustration of Precession Orbit}

\begin{figure}[h!]
    \centering
    \includegraphics[width=0.8\textwidth]{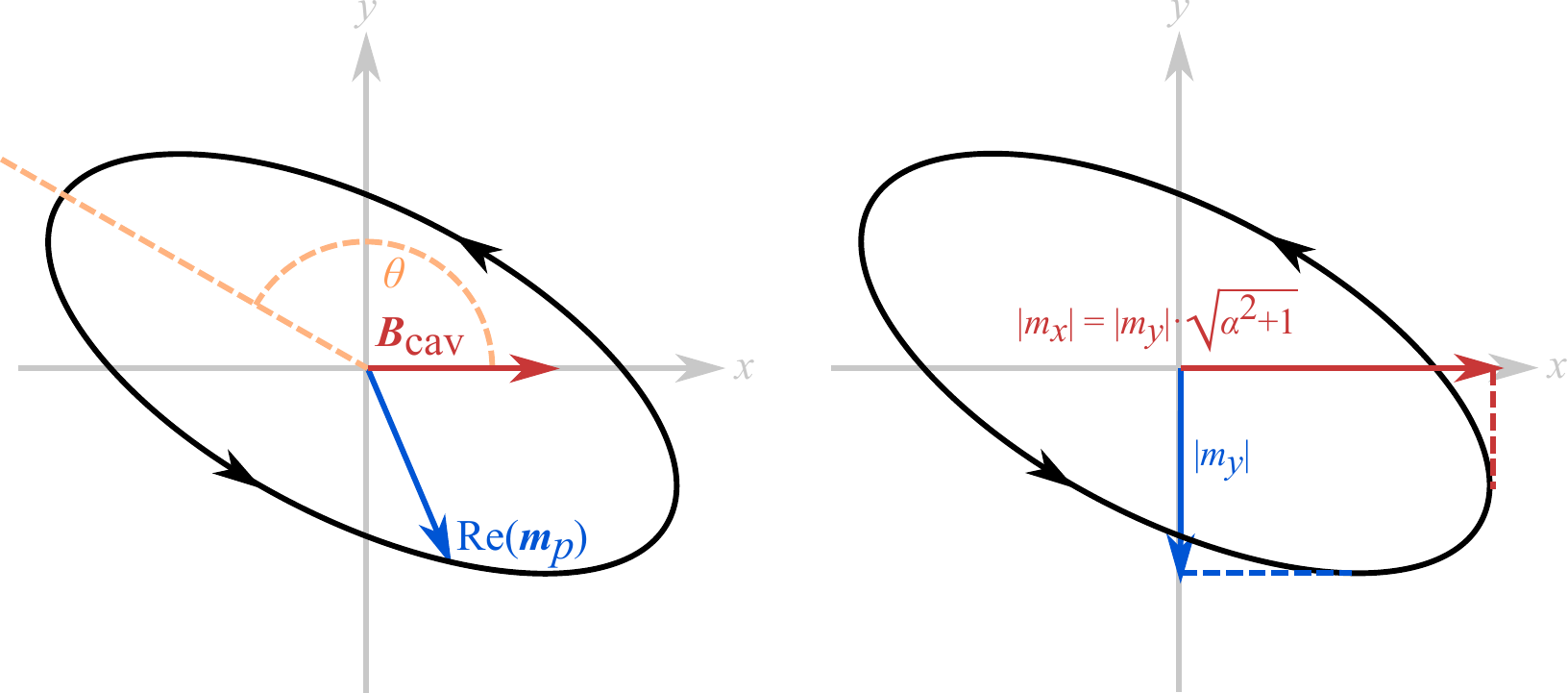}
    \caption{\textbf{Left}: $\theta$ is the angle between the $x$-axis and major axis of the elliptic orbit. $\mathrm{Re}(\pmb{m}_{p})$ is (the real part of) the precessional component of the magnetization vector of the FM. $\pmb{B}_{\mathrm{cav}}$ is the cavity magnetic field, oscillating along the $x$-axis. \textbf{Right}: The magnitudes of $m_x$ and $m_y$, in particular their relative sizes. $\alpha$ is the damping parameter of the LLG equation.}
    \label{fig:refereeA1}
\end{figure}

\end{document}